\documentclass[journal,twoside,web]{ieeecolor}
\usepackage{tmi}
\usepackage{booktabs}
\usepackage{cite}
\usepackage{amsmath}
\usepackage{amsfonts}
\usepackage{algorithmic}
\usepackage{dblfloatfix}
\usepackage{graphicx}
\usepackage{textcomp}
\usepackage{caption}
\usepackage{subcaption}
\usepackage{atbegshi}
\usepackage{xstring}
\usepackage{flushend}
\usepackage[normalem]{ulem}
\usepackage{tikz}
\usepackage{multirow}
\usepackage{color}
\definecolor{forestgreen}{RGB}{0,139,69}
\usepackage{xcolor}
\definecolor{citecolor}{HTML}{0071bc}

\usepackage{xcolor}
\definecolor{SeaGreen4}{RGB}{0,205,102} 
\definecolor{SlateBlue}{RGB}{106,90,205} 
\definecolor{DarkRed}{RGB}{178,34,34} 

\usepackage[colorlinks, linkcolor=red,  anchorcolor=DarkRed, citecolor=SeaGreen4]{hyperref} 

\usepackage{makecell}
\usepackage{amssymb}
\usepackage{CJKutf8} 

\definecolor{darkred}{rgb}{.8,.0,.0}
\newcommand\copyrighttext{%
    \footnotesize \textcopyright 2024 IEEE. Personal use of this material is permitted. Permission from IEEE must be obtained for all other uses, including republication/redistribution.}
\newcommand\copyrightnotice{%
\begin{tikzpicture}[remember picture,overlay]
\node[anchor=south,yshift=10pt] at (current page.south) {\fbox{\parbox{\dimexpr\textwidth-\fboxsep-\fboxrule\relax}{\copyrighttext}}};
\end{tikzpicture}%
}

\AtBeginDocument{\AtBeginShipoutNext{\AtBeginShipoutDiscard}}
\hypersetup{
    colorlinks=true,
    linkcolor=blue,
    filecolor=magenta,      
    urlcolor=cyan,
    pdftitle={Coronary Plaque Meshes},
    }
    
\def\BibTeX{{\rm B\kern-.05em{\sc i\kern-.025em b}\kern-.08em
    T\kern-.1667em\lower.7ex\hbox{E}\kern-.125emX}}
\begin{document}

\title{ Pre-training on High Definition X-ray Images: An Experimental Study}  



\author{Xiao Wang, Yuehang Li, Wentao Wu, Jiandong Jin, Yao Rong, Bo Jiang, Chuanfu Li, Jin Tang \\ } 

\thanks{$\bullet$ Xiao Wang, Yuehang Li, Bo Jiang, and Jin Tang are with Information Materials and Intelligent Sensing Laboratory of Anhui Province, Anhui University, Hefei 230601, China; Anhui Provincial Key Laboratory of Multimodal Cognitive Computation, Anhui University, Hefei 230601, China; School of Computer Science and Technology, Anhui University, Hefei 230601, China. (email: \{xiaowang, tangjin\}@ahu.edu.cn)} 

\thanks{$\bullet$ Wentao Wu, Jiandong Jin are with the School of Artificial Intelligence, Anhui University, Hefei 230601, China.} 

\thanks{$\bullet$ Chuanfu Li is with the First Affiliated Hospital of Anhui University of Chinese Medicine, Hefei 230022, China (email: licf@ahtcm.edu.cn).} 

\thanks{$\bullet$ Corresponding author: Bo Jiang (email: jiangbo@ahu.edu.cn)}

\maketitle
\copyrightnotice

\begin{abstract}
Existing X-ray based pre-trained vision models are usually conducted on a relatively small-scale dataset (less than 500k samples) with limited resolution (e.g., 224 $\times$ 224). However, the key to the success of self-supervised pre-training large models lies in massive training data, and maintaining high resolution in the field of X-ray images is the guarantee of effective solutions to difficult miscellaneous diseases. In this paper, we address these issues by proposing the first high-definition (1280 $\times$ 1280) X-ray based pre-trained foundation vision model on our newly collected large-scale dataset which contains more than 1 million X-ray images. Our model follows the masked auto-encoder framework which takes the tokens after mask processing (with a high rate) is used as input, and the masked image patches are reconstructed by the Transformer encoder-decoder network. 
More importantly, we introduce a novel context-aware masking strategy that utilizes the chest contour as a boundary for adaptive masking operations. 
We validate the effectiveness of our model on two downstream tasks, including X-ray report generation and disease recognition. Extensive experiments demonstrate that our pre-trained medical foundation vision model achieves comparable or even new state-of-the-art performance on downstream benchmark datasets. 
The source code and pre-trained models of this paper will be released on \hyperlink{https://github.com/Event-AHU/Medical_Image_Analysis}{https://github.com/Event-AHU/Medical\_Image\_Analysis}. 
\end{abstract}

\begin{IEEEkeywords}
High Definition X-ray Image, Pre-trained Big Models, Masked Auto Encoder, Medical Report Generation
\end{IEEEkeywords}

\section{Introduction} \label{sec:introduction}

\IEEEPARstart{M}{edical} image analysis based on X-ray is one of the most important research directions in smart healthcare. The common applications of X-ray include lesion segmentation~\cite{Ronneberger_2015_U-Net}, detection~\cite{Yan_2018_DeepLesion}, disease prediction~\cite{Rajpurkar_2017_CheXNet}, and medical report generation~\cite{Xu_2015_Show}. Previous works usually focus on a single task based on pre-trained backbone networks on ImageNet dataset~\cite{Deng_2009_ImageNet} which are deep convolutional neural networks like VGG~\cite{Simonyan_2015_Very}, ResNet~\cite{He_2016_Deep}, etc. Early deep learning methods greatly accelerated the development of medical image analysis but gradually reached their performance bottleneck. Potential reasons include privacy issues with medical data leading to scarcity, experienced doctors being unable to provide high-quality annotated data due to various reasons, limitations of the receptive field of CNN models, etc.

Inspired by the success of self-attention based Transformer~\cite{Vaswani_2017_Attention} and self-supervised learning~\cite{Chen_2020_Simple} in the natural language processing community, the researchers also designed new architectures for the perceptron of image/video data. Specifically, the ViT~\cite{Dosovitskiy_2021_Image} and Swin-Transformer network~\cite{Liu_2021_Swin} stand out as top contenders, the self-supervised learning strategies such as reconstruction-based masked auto-encoder~\cite{he2022MAE}, and contrastive learning~\cite{Radford_2021_Learning}, all sparking a new wave of research in the academic community. Naturally, these methods and techniques have also been introduced into the field of X-ray image analysis, and some progress and achievements have been made. To be specific, 
Wu et al. propose MedKLIP~\cite{wu2023medklip} which uses the paired image-text reports (about 227k studies from the MIMIC-CXR v2 dataset) for domain-specific knowledge extraction to enhance the medical image-language pre-training. 
Chen et al. propose to bridge the fusion-encoder and dual-encoder type for medical vision-text pre-training via PTUnifier~\cite{PTUnifier2023}. 
Multi-Modal Masked auto-encoder ($M^3AE$) is proposed by Chen et al.~\cite{chen2022M3ME} which attempt to learn cross-modal domain knowledge in a self-supervised learning manner by reconstructing missing pixels and tokens from randomly masked images and texts. 
Xiao et al.~\cite{xiao2023delvingMAE} verified that the pre-training of ViT on 266,340 chest X-rays using MAE can achieve better results than CNN model DenseNet-121 using the MoCo v2 framework.  
However, it is easy to find that the challenging X-ray image based tasks are still far from being solved well.

According to our observation, reflection, and discussions and consultations with senior experts in the medical field, we believe that these models may still be limited by the following factors: 
\begin{itemize}
    \item \textbf{The conflict between the high resolution of X-ray images and the standard resolution of pre-trained models used in natural images}: Specifically, existing models are typically trained on standard image resolutions, such as $224 \times 224$, while actual X-ray image resolutions may reach levels as high as $2000 \times 3000$. This discrepancy can lead to a significant loss of image information originally present in high-resolution data when down-sampling occurs. 
    \item \textbf{Existing works rarely take into account the contextual prior information of chest X-ray images}: Standard MAE pre-trained models utilize random sampling strategies for token masking, with some works~\cite{he2022MAE} considering the impact of different mask rates on the results, while overlooking the importance of differences between the inside and outside of the chest contour.
    \item \textbf{Current X-ray based large-scale models have only been pre-trained on small-scale datasets.} Existing X-ray image based pre-trained models are usually pre-trained on public datasets, such as MIMIC~\cite{johnson2019mimicCXR} which contains about 300K images only. This level of data might not be abundant for large-scale model pre-training. 
\end{itemize}
Considering the aforementioned issues, it is natural to raise the following questions: ``\textit{Can we further expand the scale of X-ray images for the pre-training task? Is it possible to conduct this pre-training task at higher resolutions? Could more prior information be leveraged to improve the effectiveness of pre-training?}"

In this paper, we answer these questions by proposing a novel MAE framework that adopts the context-aware masking strategy pre-trained on massive (about 1 million) high-definition X-ray images. As shown in Fig.~\ref{fig:framework}, we resize the input X-ray image into $1280 \times 1280$ and believe the high-definition images can better preserve the detailed information of the original data. Then, we partition it into non-overlapping image patches and project the patches into token representations. The input tokens are dropped out with a high ratio (larger than $70\%$) by following the MAE~\cite{he2022MAE}, however, we propose a novel context-aware masking strategy instead of random masking used in MAE. Because the chest X-ray images contain prominent contour line information, and typically, doctors are more concerned with information about lesions in the chest area. The visible tokens are added with position encodings, then, we feed them into a Transformer encoder. The masked tokens are randomly initialized and concatenated with the output of the Transformer encoder and fed into the Transformer decoder network for masked region reconstruction. After the pre-training phase is finished, we extract the Transformer encoder as the backbone network for downstream tasks to validate its effectiveness in these tasks, including Chinese/English report generation and disease prediction.

To sum up, the contributions of this paper can be summarized as the following three aspects: 

$\bullet$ We propose the first pre-trained foundation model using high-definition X-ray images ($1280 \times 1280$), based on the masked auto-encoder framework. 

$\bullet$ We exploit a new context-aware masking strategy for the X-ray image based masked auto-encoder framework. 

$\bullet$ We conduct extensive experiments on multiple downstream tasks, including Chinese/English medical report generation and disease classification.

\textbf{The rest of this paper is organized as follows}: 
we first introduce the related works in section~\ref{sec:relatedworks} by reviewing pre-training on the medical images, and downstream tasks. In section~\ref{sec:Method}, we focus on describing our proposed framework, including an overview, pre-training stage, and downstream tasks. In section~\ref{sec:experiments}, we conduct extensive experiments to validate the effectiveness of our model from both qualitative and quantitative views. Finally, we conclude this paper and propose future works in section~\ref{sec:conclusion}.

\section{Related Works} \label{sec:relatedworks} 

In this section, we will introduce the algorithms mostly related to ours, including Pre-training on Medical Images and Downstream Tasks. More works can be found in the following surveys~\cite{wang2023MMPTMs, shrestha2023medicalVLPsurvey, azad2023foundational, zhao2023clipMISurvey, thirunavukarasu2023large}.

\subsection{Pre-training on Medical Image} 
The pre-training techniques proposed for the medical image analysis can be categorized into two main streams, i.e., the \textit{contrastive learning} based~\cite{huang2021gloria, liu2023g2d, zhan2023unidcp, wang2023PhenotypeCLIP, wu2023medklip} and \textit{masked token based reconstruction} schemes~\cite{zhang2023MPMA, chen2022M3ME, zhou2022MRM}. A brief summary of these models is provided in Table~\ref{tab:pretrainedMedicalModels}. 
For the masked token based reconstruction frameworks, the inputs are usually masked with a high ratio and attempt to reconstruct them using an encoder-decoder network. Specifically, Chen et al.~\cite{chen2022M3ME} propose the $M^3AE$ which takes the masked medical image and language as the input and conducts feature-level fusion using cross-attention. They propose two independent decoders for the reconstruction of vision and language modality. Zhou et al.~\cite{zhou2022MRM} introduce the MRM (masked record modeling) framework which reconstructs masked image patches and masked report tokens following a multi-task scheme. Xiao et al.~\cite{xiao2023delvingMAE} pre-train a foundation model based on masked autoencoder~\cite{he2022MAE} and conduct extensive ablation studies on the advantages between ViT and CNN.  

For the contrastive learning based models, the relations between the medical images and reports are mainly considered for pre-training. Specifically, GLoRIA~\cite{huang2021gloria} proposed by Huang et al. which is an attention-based framework for learning global and local representations by contrasting image sub-regions and words in the paired report. G2D~\cite{liu2023g2d} (Global to Dense level representation learning) proposed by Liu et al. is a medical vision-language pre-training framework that improves the granularity and more accurate grounding for the learned features. Zhan et al. propose the UniDCP~\cite{zhan2023unidcp} which is a unified model and supports multiple medical fine-tuning tasks. They design cross-modal prompts to harmonize heterogeneous inputs from multiple pre-training tasks. 
Wang et al. propose the PhenotypeCLIP~\cite{wang2023PhenotypeCLIP} which also follows the contrastive learning framework and learns more fine-grained phenotype-based representations to bridge the gap between vision and language efficiently. CXR-CLIP~\cite{you2023cxrCLIP} first generates image-text pairs from image-label datasets via prompt engineering and conducts pre-training using three kinds of contrastive losses. 
Different from these works which conduct pre-training on low-resolution X-ray images or contrastive learning between X-ray and English text, in this work, we propose a pre-trained foundation model on high-definition X-ray images and support medical report generation in both English and Chinese.

\begin{table*}
\centering
\resizebox{\textwidth}{!}{ 
\begin{tabular}{c|l|l|l|c|c|c|c|c}
\hline \toprule [0.5 pt]
\textbf{No.} & \textbf{Name} & \textbf{Publication} & \textbf{Pre-train Data} & \textbf{Backbone} & \textbf{Modality} & \textbf{\begin{tabular}[c]{@{}c@{}}Pre-train\\ paradigm\end{tabular}} & \textbf{\begin{tabular}[c]{@{}c@{}}Downstream\\ Tasks\end{tabular}} & \textbf{URL} \\ \hline
01 & ARL~\cite{chen2022align} & ACMMM-2022 & \begin{tabular}[c]{@{}l@{}}ROCO, MedICaT,\\ MIMIC-CXR 771k($224\times224$)\end{tabular} & \begin{tabular}[c]{@{}c@{}}CLIP-ViT-B\\ + RoBERTa-base\end{tabular} & Image-Text & ARL & \begin{tabular}[c]{@{}c@{}}VQA, DP,\\ retrieval\end{tabular} & \hyperlink{https://github.com/zhjohnchan/ARL}{GitHub} \\ \hline
02 & $M^3AE$~\cite{chen2022M3ME} & MICCAI-2022 & \begin{tabular}[c]{@{}l@{}}ROCO,\\ MedICaT 298k($256\times256$)\end{tabular} & \begin{tabular}[c]{@{}c@{}}CLIP-ViT-B\\ + RoBERTa-base [\end{tabular} & Image-Text & $M^3AE$ & VQA, DP, retrieval & \hyperlink{https://github.com/zhjohnchan/M3AE}{GitHub} \\ \hline
03 & MedKLIP~\cite{wu2023medklip} & ICCV-2023 & MIMIC-CXR  377k($224\times224$) & \begin{tabular}[c]{@{}c@{}}ResNet-50\\ + ClinicalBERT\end{tabular} & Image-Text & KLIP & DP, Seg & \hyperlink{https://github.com/MediaBrain-SJTU/MedKLIP}{GitHub} \\ \hline
04 & Medical\_MAE~\cite{xiao2023delvingMAE} & WACV-2023 & \begin{tabular}[c]{@{}l@{}}ChestX-Ray14, CheXpert, \\ MIMIC-CXR ($256\times256$)\end{tabular} & ViT-S & Image & MAE & DP & \hyperlink{https://github.com/lambert-x/Medical_MAE}{GitHub} \\ \hline
05 & ECAMP~\cite{wang2023ecamp} & arXiv-2023 & MIMIC-CXR 377k($448\times448$) & ViT & Image-Text & MAE & DP & \hyperlink{https://github.com/ToniChopp/ECAMP}{GitHub} \\ \hline
06 & MRM~\cite{zhou2023advancing} & ICLR-2023 & MIMIC-CXR 377k($224\times224$) & ViT & Image-Text & MAE & DP & \hyperlink{https://github.com/RL4M/MRM-pytorch}{GitHub} \\ \hline
07 & G2D~\cite{liu2023g2d} & arXiv-2023 & MIMIC-CXR 213k($256\times256$) & \begin{tabular}[c]{@{}c@{}}ResNet-50\\ + ClinicalBERT\end{tabular} & Image-Text & G2D & DP, Seg, Detect & - \\ \hline
08 & UniDCP~\cite{zhan2023unidcp} & arXiv-2023 & \begin{tabular}[c]{@{}l@{}}ROCO,\\ MIMIC-CXR 458k($224\times224$)\end{tabular} & CLIPViT & Image-Text & UniDCP & \begin{tabular}[c]{@{}c@{}}VQA, RG, DP,\\ Seg, retrieval\end{tabular} & - \\ \hline
09 & T3D~\cite{liu2023t3d} & arXiv-2023 & BIMCV 8k( $96\times96\times96$) & \begin{tabular}[c]{@{}c@{}}SwinUNTER\\ + RadBERT\end{tabular} & 3D Volume-Text & T3D & DP, Seg & - \\ \hline
10 & PhenotypeCLIP~\cite{wang2023PhenotypeCLIP} & ACL-2023 & CheXpert & ResNet-50 + BERT & Image-Text & PhenotypeCLIP & RG & - \\ \hline
11 & MaCo~\cite{huang2023enhancing} & arXiv-2023 & MIMIC-CXR & ViT-B + BERT & Image-Text & MAE & DP, Seg & - \\ \hline
12 & CXR-CLIP~\cite{you2023cxrCLIP} & MICCAI-2023 & \begin{tabular}[c]{@{}l@{}}MIMIC-CXR, CheXpert,\\ ChestX-ray14 528k($224\times224$)\end{tabular} & ResNet-50 & Image-Text & CLIP & DP, retrieval & \hyperlink{https://github.com/kakaobrain/cxr-clip}{GitHub} \\ \hline
13 & PTUnifier~\cite{chen2023towards} & ICCV-2023 & \begin{tabular}[c]{@{}l@{}}ROCO, MedICaT, MIMIC-CXR \\ 437k($288\times288$)\end{tabular} & \begin{tabular}[c]{@{}c@{}}CLIP-ViT-B \\ + RoBERTa-base\end{tabular} & Image-Text & PTUnifier & \begin{tabular}[c]{@{}c@{}}VQA, RG, DP,\\ retrieval\end{tabular} & \hyperlink{https://github.com/zhjohnchan/ptunifier}{GitHub} \\ \hline
14 & MPMA~\cite{zhang2023MPMA} & TMM-2023 & \begin{tabular}[c]{@{}l@{}}ROCO,\\ MIMIC-CXR 458k($224\times224$)\end{tabular} & ViT + BERT & Image-Text & MPMA & DP, RG, VQA & - \\ \hline
15 & IMITATE~\cite{liu2023imitate} & arXiv-2023 & MIMIC-CXR 377k($224\times224$) & \begin{tabular}[c]{@{}c@{}}ResNet-50\\ + Bio-ClinicalBERT\end{tabular} & Image-Text & IMITATE & DP, Seg, Detect & - \\ \hline
16 & MeDSLIP~\cite{fan2024medslip} & arXiv-2024 & MIMIC-CXR 377k($224\times224$) & \begin{tabular}[c]{@{}c@{}}ResNet-50\\ + Bio-ClinicalBERT\end{tabular} & Image-Text & ProtoCL & DP, Seg, & - \\ \hline
17 & ASG~\cite{li2024anatomical} & arXiv-2024 & MIMIC-CXR 377k($224\times224$) & \begin{tabular}[c]{@{}c@{}}ResNet50 / ViT-B\\ + BioClinicalBERT\end{tabular} & Image-Text & ASG & DP, Seg & - \\ \hline
18 & MLIP~\cite{liu2024mlip} & arXiv-2024 & MIMIC-CXR 377k($224\times224$) & \begin{tabular}[c]{@{}c@{}}ViT-B\\ + BioClinicalBERT\end{tabular} & Image-Text & MLIP & DP, Seg, & - \\ \hline
 & Ours & - & \textbf{1M ($1280\times1280$)} & ViT & Image &MAE  & RG, DP & \hyperlink{https://github.com/Event-AHU/Medical_Image_Analysis}{GitHub} \\ 
\hline \toprule [0.5 pt]
\end{tabular}} 
\caption{Comparison between our model and existing X-ray based pre-trained foundation models.
            RG and DP are short for Report Generation and Disease Prediction, respectively.} 
\label{tab:pretrainedMedicalModels}
\end{table*}

\subsection{Downstream Tasks} 
In this paper, we focus on handling two representative tasks, including \textit{medical report generation}~\cite{hyland2023MAIRA, chiang2023vicuna, li2019knowledge, hou2023organ, liu2021auto, wang2022cross, zhang2023semi, jin2023promptmrg} and \textit{disease classification}~\cite{Li_2023_ICCV, cheng2023prior, chen2023bomd, tanida2023interactive, huang2023kiut, wang2023metransformer}. 
For the report generation, Stephanie et al. propose the MAIRA-1~\cite{hyland2023MAIRA} which combines CXR-specific image encoder and fine-tuned LLM based on Vicuna-7B~\cite{chiang2023vicuna}. Li et al.~\cite{li2019knowledge} propose a Knowledge-driven Encode, Retrieve, Paraphrase (KERP) method. At its core is the proposed universal implementation unit-Graph Transformer (GTR), which dynamically transforms high-level semantics across various domains of graph-structured data, including knowledge graphs, images and sequences. ORGAN~\cite{hou2023organ} proposed by Hou et al. is an  observation-guided radiology report generation framework. It first generates an observation plan, feeds both the plan and images into the report generation process, and uses an observation graph and a tree-based reasoning mechanism to accurately enrich the information of each observation by capturing multiple formats. Liu et al.~\cite{liu2021auto} proposed an unsupervised model called Knowledge Graph Auto-Encoder (KGAE), which comprises a pre-constructed knowledge graph, a knowledge-driven encoder and a knowledge-driven decoder. In the absence of paired image-report training data, unsupervised KGAE can generate desirable medical reports. Wang et al.~\cite{wang2022cross} explore cross-modal feature interactions and propose a Cross-modal PROtotype driven NETwork (XPRONET) to promote cross-modal pattern learning and leverage it to enhance the task of radiology report generation. Zhang et al.~\cite{zhang2023semi} utilizes the Graph-guided Hybrid Feature Encoding (GHFE) module to encode the intrinsic relationships between pathological changes into graph embeddings using prior disease knowledge graphs. The GHFE combines graph embeddings, semantic embeddings, and visual features to form hybrid features, which are then fed into the decoder of a Transformer to generate reports. PromptMRG~\cite{jin2023promptmrg} proposed by Jin et al. which converts the diagnostic results of disease classification branches into prompts to guide report generation. By utilizing cross-modal retrieval and dynamic feature aggregation, it further enhances diagnostic accuracy. Different from existing medical report generation works, we propose a novel context-aware masking strategy for Chinese/English medical report generation.

For the disease classification, Li et al.~\cite{Li_2023_ICCV} propose the Unify, Align, and Refine (UAR) method, which introduces the Latent Space Unifier, Cross-modal Representation Aligner and Text-to-Image Refiner to learn multi-level cross-modal alignments. The authors of~\cite{cheng2023prior} present a prototype representation learning framework that integrates global and local alignment between medical images and reports. By constructing a sentence-wise prototype memory bank, the network can focus on low-level local visual features and high-level clinical language features. 
BoMD~\cite{chen2023bomd} proposed by Chen et al. is a method designed for learning noisy multi-label CXR by detecting and re-labeling noisy samples from the dataset in a smooth manner. It optimizes a set of multi-label descriptors to promote their similarity with the semantic descriptors generated by a multi-label image annotation language model. 
Tanida et al.~\cite{tanida2023interactive} propose a method that focuses explicitly on highlighted anatomical regions through object detection and generates descriptions for specific regions. Its interactive functionality allows radiologists to directly participate in the decision-making process. 
Kiut~\cite{huang2023kiut} proposed by Huang et al. is designed to learn multi-level visual representations. It incorporates a u-connection schema to simulate interactions between different modalities and has developed a symptom graph and an injectable knowledge distiller to assist in report generation. 
Wang et al.~\cite{wang2023metransformer} introduced multiple learnable \textit{expert} tokens to encourage these expert tokens to capture complementary information through orthogonal loss. Each participating expert token guides the cross-modal attention between input words and visual tokens, enhancing the quality of generated reports. 
In contrast to existing disease prediction approaches, we employ a foundation model pre-trained on high-resolution X-ray images and have developed a novel context-aware masking strategy based on an X-ray image masked auto-encoder framework.

\section{Our Proposed Approach} \label{sec:Method}

In this section, we will first give an overview to help better understand our pre-training framework. Then, we dive into the details of pre-training on high-definition X-ray images. After that, we will introduce the downstream tasks used to validate the effectiveness of our pre-trained model, including Chinese/English report generation and disease prediction.

\begin{figure*}
    \centering
    \includegraphics[width=1\linewidth]{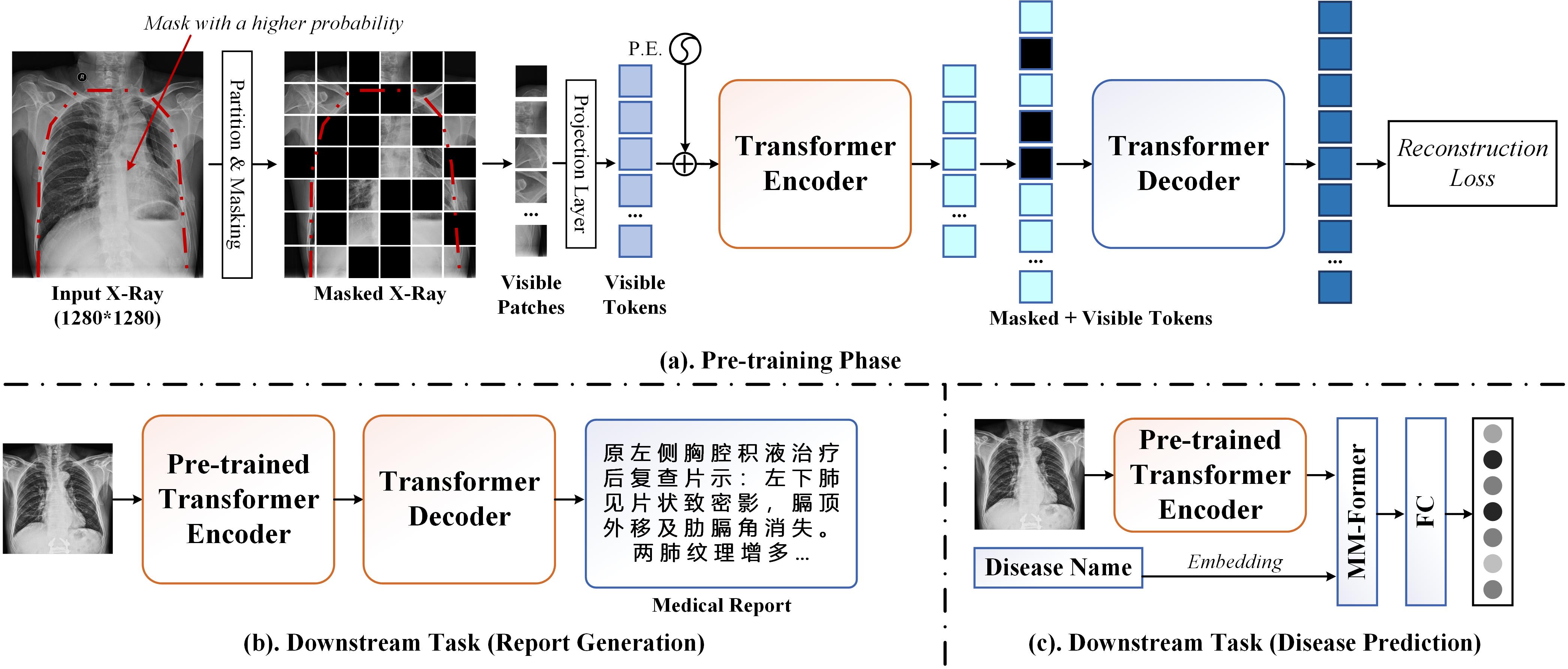}
    \caption{ (a) An illustration of our proposed high-definition X-ray image based pre-training framework using masked auto-encoder. (b, c) are two downstream tasks used for the validation of our pre-training framework. } 
    \label{fig:framework}
\end{figure*}

\subsection{Overview} 
As illustrated in Fig.~\ref{fig:framework}, our pre-training scheme follows the masked auto-encoder framework~\cite{he2022MAE} that attempts to reconstruct the highly masked input patches. Note that, we take the high-definition X-ray images as the input to better retain its raw detailed information. The image is partitioned into non-overlapping regions and transformed into token representations using a convolutional layer. Inspired by the fact that the cue in the chest part may be more important than other regions, therefore, we introduce a simple but effective context-aware masking strategy by masking more patches inside the chest regions. We believe this will help the model to focus on these regions in the pre-training phase. The visible tokens are fed into the ViT encoder and the outputs are concatenated with the randomly initialized masked tokens. Then, a Transformer decoder network is adopted to reconstruct the input image. Once the pre-training is finished, we fine-tuning the Transformer encoder for the downstream task to achieve a higher performance. More details will be introduced in subsequent sub-sections, respectively.

\subsection{Pre-training Stage} 

In this section, we will focus on the details of our pre-training from the perspective of Input Processing, Context-Aware Masking, Transformer Encoder and Decoder, and Loss Function.

\noindent \textbf{Input Processing. }  
Given the raw X-ray images, we first resize them into a fixed-resolution $\mathcal{I} \in \mathbb{R}^{H \times W \times 1}$, here, we set both $H$ and $W$ as 1280. Following the Transformer encoder used in MAE~\cite{he2022MAE}, we partition the whole image into $N$ non-overlapping regions $\{P_1, P_2, ..., P_N\}$, and the resolution of each region is $64 \times 64 \times 1$. Then, we adopt a convolution layer (kernel size $64 \times 64$) to transform the image patches into the token representations $\{X_1, X_2, ..., X_N\}$ whose dimension is 1024. 
After we get these tokens, the MAE framework masks them with a high ratio, and the rest of them are treated as visible tokens. For example, He et al.~\cite{he2022MAE} mask $70\%$ of them for the natural image, and Xiao et al.~\cite{xiao2023delvingMAE} claim that the best downstream performance can be achieved when removing $90\%$ of them in the pre-training stage of their MAE-based X-ray model. However, seldom of they consider the context information of chest X-ray images for the masking operation.

\noindent \textbf{Context-Aware Masking. }  
In this work, we propose a novel context-aware masking strategy to process the transformed tokens. The key insight is that more useful cues can be mined in the chest region of X-ray image. Specifically, we manually define a boundary line of the chest, as shown in Fig.~\ref{fig:framework}, and mask the tokens inside of the chest line with a higher probability. 
Once we remove the masked tokens, the visible tokens $X_i, i \in \{1, 2, ..., M\}$ are added with position encodings $E_i, i \in \{1, 2, ..., M\}$, therefore, we feed the $\mathcal{X}_i = X_i + E_i, i \in \{1, 2, ..., M\}$ into the Transformer encoder network.

\noindent \textbf{Transformer Encoder and Decoder. }  
In our practical implementation, we adopt the ViT-L~\cite{Dosovitskiy_2021_Image} (16 heads and 1024 embedding dimension, 304M trainable parameters) as the Transformer encoder network which contains 24 Transformer blocks. The key operator in the Transformer is multi-head self-attention and the detailed computing process of self-attention can be formulated as: 
\begin{equation}
    \label{MHSA} 
    SelfAttenion = Softmax(\frac{QK^{T}}{\sqrt{c}}) V 
\end{equation}
where $Q, K$ and $V$ are processed input tokens $\mathcal{X}_i$, $c$ is the dimension of input tokens. $Softmax(\cdot)$ denotes the Softmax layer.

\begin{figure}
    \centering
    \includegraphics[width=0.8\linewidth]{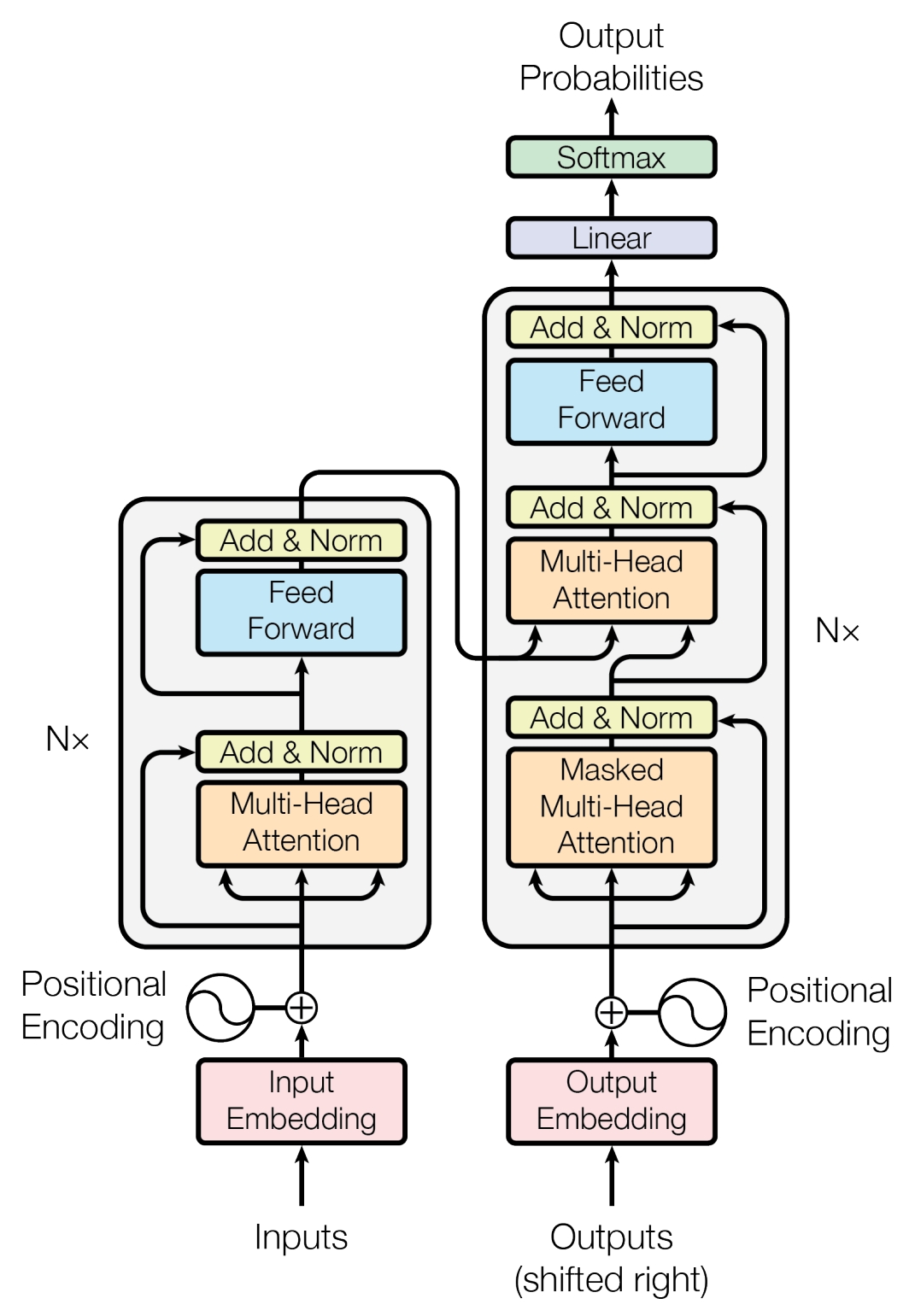}
    \caption{The detailed architectures of Transformer from \cite{Vaswani_2017_Attention}.}
    \label{fig:transformer}
\end{figure}

Given the output of Transformer encoder, we integrate them with the masked tokens, whose parameters are randomly initialized, and feed into the Transformer decoder network. For the detailed network architecture of the Transformer decoder, we directly borrow from the vanilla MAE framework for the masked X-ray image reconstruction. It contains 8 Transformer blocks.

\noindent \textbf{Loss Function. }  
After we obtain the reconstructed image patches, we compute its distance with the ground truth image patch using the $L_2$ loss function. Note that, existing work~\cite{xiao2023delvingMAE} demonstrates that other loss functions like $L_1$, \textit{smooth}-$L_1$, \textit{SSIM}, and \textit{adversarial loss} do not improve the MAE framework.

\subsection{Downstream Tasks}  
In this work, we validate our proposed framework by introducing the pre-trained Transformer encoder (i.e., ViT-L) into two downstream tasks, including X-ray based report generation and disease prediction, as illustrated in Fig.~\ref{fig:framework} (b) and (c).

\noindent \textbf{X-ray based Report Generation. }  
Given the X-ray image, the task of report generation targets describing the disease information using natural language. Usually, this task is formulated as an English sentence generation. In addition, we also build a new X-ray dataset for Chinese report generation. More details about this dataset will be introduced in section \ref{downstreamDATA}. In our implementation, we build our report generator based on the R2Gen\footnote{\url{https://github.com/zhjohnchan/R2Gen}} toolkit proposed in~\cite{chen2020R2Gen}.

\noindent \textbf{Disease Prediction. }  
This task can be treated as a standard multi-category classification problem by mapping the input X-ray image into a distribution of the response score of each category. However, this may ignore the semantic information of category names which is also useful for high-performance recognition. In this work, we follow the VTB\footnote{\url{https://github.com/cxh0519/VTB}}~\cite{cheng2022VTB}, which formulates the multi-label classification task as a vision-text fusion problem, for the disease prediction. As shown in Fig.~\ref{fig:framework} (c), we adopt the pre-trained ViT encoder to extract the features of the input X-ray image and utilize the CLIP text encoder to embed the given disease name. Then, we fuse the two modalities using a multi-modal Transformer network and predict the disease using a fully connected (FC) layer.

\section{Experiments} \label{sec:experiments}

In this section, we will first introduce the datasets, evaluation metrics, and implementation details in sub-section~\ref{datasets}, \ref{metrics}, \ref{implementDetails}, respectively. 
Then, we will focus on reporting and analyzing the results of the medical report generation and disease prediction, in sub-section~\ref{resultsMRG}, \ref{resultsDisR}. After that, we will give extensive ablation studies of our model in sub-section~\ref{ablationStudy} and visualize the reconstruction on the masked tokens, similar matric, generated medical reports, and disease predictions in sub-section~\ref{visualization}. Also, we describe the limitations of this work in sub-section~\ref{limitation}.

\subsection{Datasets} \label{datasets}

\subsubsection{Pre-training Dataset} 
\begin{figure*}  
    \centering  
    \resizebox{\textwidth}{!}{
    \begin{subfigure}[b]{0.15\textwidth}  
        \centering  
        \includegraphics[width=\textwidth]{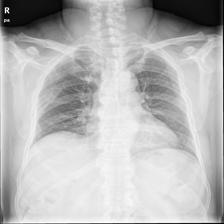}  
    \end{subfigure}  
    \begin{subfigure}[b]{0.15\textwidth}  
        \centering  
        \includegraphics[width=\textwidth]{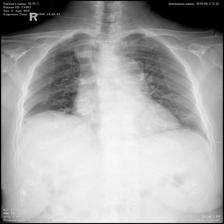}  
    \end{subfigure}  
    \begin{subfigure}[b]{0.15\textwidth}  
        \centering  
        \includegraphics[width=\textwidth]{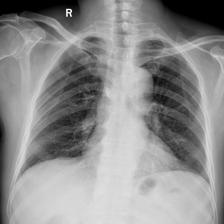}  
    \end{subfigure}  
    \begin{subfigure}[b]{0.15\textwidth}  
        \centering  
        \includegraphics[width=\textwidth]{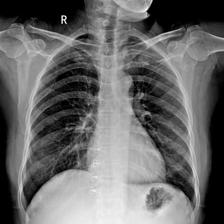}  
    \end{subfigure}  
    \begin{subfigure}[b]{0.15\textwidth}  
        \centering  
        \includegraphics[width=\textwidth]{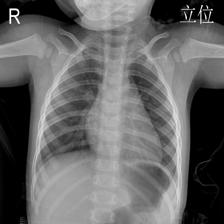}  
    \end{subfigure}  
    \begin{subfigure}[b]{0.15\textwidth}  
        \centering  
        \includegraphics[width=\textwidth]{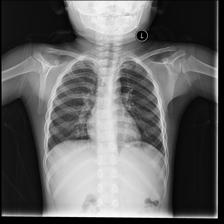}  
    \end{subfigure}  
    }
    \resizebox{\textwidth}{!}{
    \begin{subfigure}[b]{0.15\textwidth}  
        \centering  
        \includegraphics[width=\textwidth]{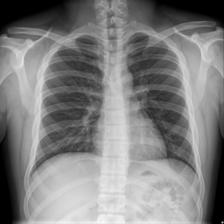}  
    \end{subfigure}  
    \begin{subfigure}[b]{0.15\textwidth}  
        \centering  
        \includegraphics[width=\textwidth]{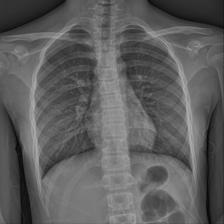}  
    \end{subfigure}  
    \begin{subfigure}[b]{0.15\textwidth}  
        \centering  
        \includegraphics[width=\textwidth]{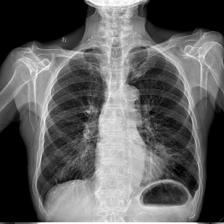}  
    \end{subfigure}  
    \begin{subfigure}[b]{0.15\textwidth}  
        \centering  
        \includegraphics[width=\textwidth]{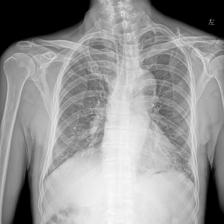}  
    \end{subfigure}  
    \begin{subfigure}[b]{0.15\textwidth}  
        \centering  
        \includegraphics[width=\textwidth]{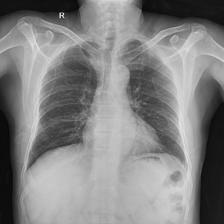}  
    \end{subfigure}  
    \begin{subfigure}[b]{0.15\textwidth}  
        \centering  
        \includegraphics[width=\textwidth]{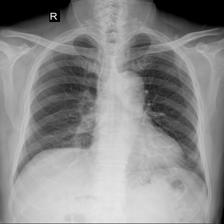}  
    \end{subfigure}  
    }
    \resizebox{\textwidth}{!}{
        \begin{subfigure}[b]{0.15\textwidth}  
        \centering  
        \includegraphics[width=\textwidth]{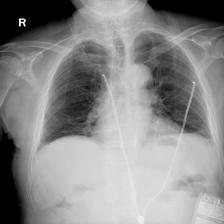}  
    \end{subfigure}  
            \begin{subfigure}[b]{0.15\textwidth}  
        \centering  
        \includegraphics[width=\textwidth]{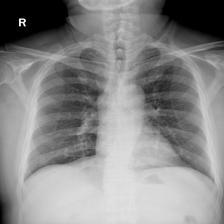}  
    \end{subfigure}
            \begin{subfigure}[b]{0.15\textwidth}  
        \centering  
        \includegraphics[width=\textwidth]{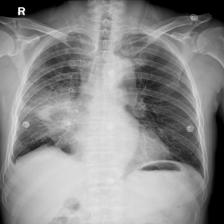}  
    \end{subfigure}
            \begin{subfigure}[b]{0.15\textwidth}  
        \centering  
        \includegraphics[width=\textwidth]{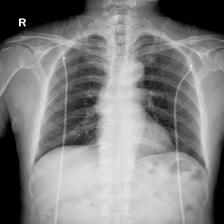}  
    \end{subfigure}
            \begin{subfigure}[b]{0.15\textwidth}  
        \centering  
        \includegraphics[width=\textwidth]{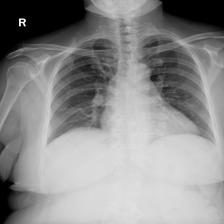}  
    \end{subfigure}
            \begin{subfigure}[b]{0.15\textwidth}  
        \centering  
        \includegraphics[width=\textwidth]{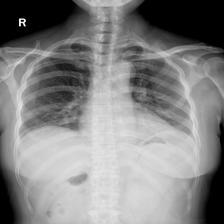}  
    \end{subfigure}
    }
    \caption{ Some representative samples of our collected PCC-Xray dataset.}  
    \label{fig:ChinesePretrainedDATA}  
\end{figure*}

To pre-train a high-performance X-ray foundation model, the first thing we need to do is the collection of large-scale X-ray images. Therefore, a large-scale and high-resolution dataset that contains $1,053,791$ X-ray medical images is collected for the pre-training. Some representative samples are visualized in Fig.~\ref{fig:ChinesePretrainedDATA}.

\subsubsection{Downstream Datasets} \label{downstreamDATA}
We conduct extensive experiments on medical report generation task and disease prediction task, and the involved datasets including \textbf{IU-Xray}~\cite{li2023dynamic}, our Private Chinese Chest X-ray image based report generation dataset (termed \textbf{PCC-Xray} in this paper), and \textbf{RSNA-Pneumonia}~\cite{shih2019RSNAPneumonia} dataset. A brief introduction to these datasets is given below.

$\bullet$ [\textit{English Report Generation}] IU-Xray dataset~\cite{li2023dynamic} is a widely used dataset for the evaluation of radiology reporting systems. It contains \textit{7,470} chest X-ray images and corresponding \textit{3,955} reports. Following previous works~\cite{li2023dynamic}, in our experiments, we utilize the processed dataset obtained by excluding samples with one image only. Specifically, our training, validation, and testing subset contains \textit{2069}/\textit{296}/\textit{590} samples, respectively.

\begin{figure}
\centering
\includegraphics[width=1\linewidth]{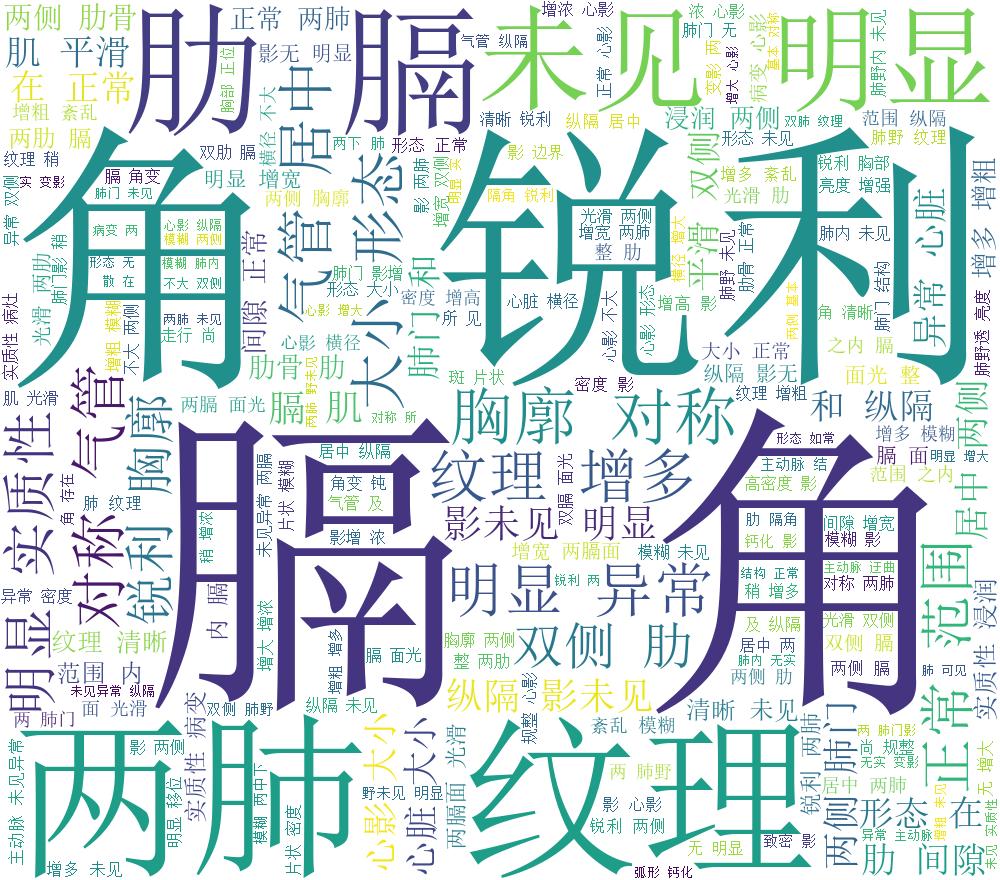}
\caption{The word cloud of our newly collected PCC-Xray dataset for Chinese medical report generation.} 
\label{fig:chinese_wordcloud}
\end{figure}

$\bullet$ [\textit{Chinese Report Generation}] PCC-Xray dataset: It was built by the First Affiliated Hospital of Anhui University of Chinese Medicine and Anhui University which contains $200,172$ high-resolution chest X-ray images. Each X-ray image is meticulously annotated with a Chinese medical report, with an average of \textit{71} Chinese characters per sentence. Regarding the Chinese medical reports of interest, we provide a word cloud as shown in Fig.~\ref{fig:chinese_wordcloud}. It can be observed that our reports cover descriptions of many common diseases and some difficult and miscellaneous conditions. We split it into the training, validation, and testing subset which contains \textit{140120}/\textit{20018}/\textit{40034} X-ray image and report pairs, respectively.

$\bullet$ [\textit{Disease Prediction}] RSNA-Pneumonia dataset~\cite{shih2019RSNAPneumonia} comprises \textit{30k} frontal view chest radiographs, each accompanied by bounding boxes indicating pneumonia opacities if present.  We follow the official data split, which includes training/validation/testing sets consisting of \textit{25184}/\textit{1500}/\textit{3000} samples, respectively.

\subsection{Evaluation Metrics} \label{metrics}
For the \textit{X-ray report generation task}, we adopt the widely used four metrics for the evaluation, including \textbf{CIDEr} (Captions Generated by Diverse Experts)~\cite{vedantam_cider_2015}, \textbf{BLEU-4} (Bilingual Evaluation Understudy-4)~\cite{papineni_bleu_2002}, \textbf{ROUGE-L} (Recall-Oriented Understudy for Gisting Evaluation - Longest Common Subsequence)~\cite{lin_rouge_2004}, and \textbf{METEOR} (Metric for Evaluation of Translation with Explicit Ordering)~\cite{banerjee_meteor_2005}. Specifically, 
CIDEr measures the consensus between the generated captions and multiple reference captions. It evaluates the quality of image captioning by computing the cosine similarity between n-grams in the generated caption and those in the reference captions. 
BLEU-4 evaluates the quality of machine-generated translations or text summaries by comparing them against reference translations or summaries. It measures the precision of n-grams (usually up to 4-grams) in the generated text compared to the reference texts. 
ROUGE-L assesses the quality of text summaries or translations by comparing them to reference texts. It focuses on the longest common subsequences between the generated and reference texts, emphasizing recall. 
METEOR evaluates machine-generated translations or summaries by considering both unigram precision and recall, as well as the alignment between the generated and reference texts. It also incorporates stemming and synonymy matching.

For the \textit{disease prediction task}, in this work, we adopt the \textbf{AUROC} (Area Under the Receiver Operating Characteristic Curve), \textbf{F1}, and \textbf{Accuracy} metrics to compare the performance of different models. To be specific, 
AUROC is a metric commonly used to evaluate binary classification models. It measures the ability of the model to distinguish between positive and negative samples across different decision thresholds. 
The F1 score is the harmonic mean of precision and recall. It's commonly used in binary classification tasks to provide a single metric that balances both precision and recall. 
Accuracy simply measures the proportion of correctly classified instances out of the total instances evaluated.

\subsection{Implementation Details} \label{implementDetails}

$\bullet$ \textbf{Pre-training Stage.~} 
In the pre-training phase, we resize the X-ray image into a fixed resolution, i.e., $1280 \times 1280$. The learning rate is set as 0.00025, and the weight decay is 0.04. The batch size is 1024 and training for a total of 83 epochs on our dataset. The AdamW~\cite{adamW} is adopted as the optimizer. 
The pre-training is conducted on a server with eight NVIDIA A800 GPUs (80GB) and about 660 hours are needed for our pre-training phase.

$\bullet$ \textbf{Downstream Tasks.~} 
For the generation of Chinese medical reports, we fine-tune the model using our PCC-Xray dataset, which comprises a total of 200,000 X-ray images. The training configuration is as follows: we resize the images to $224 \times 224$, set the batch size to 16, employ RoBERTa~\cite{liu2019roberta} as the tokenizer, and conduct training over 60 epochs. In the case of English medical report generation, we fine-tune the model on the IU-Xray dataset. Before inputting the images into the model, we resize them to $384 \times 384$. We set the batch size to 16, establish a maximum sequence length of 30, and keep other parameters identical to R2Gen. 

For the disease prediction task, we performed experiments on the RSNA-Pneumonia dataset using code derived from the Visual-Textual Baseline (VTB). We configured the batch size to 200 and set the input image size to $224 \times 224$, while leaving the remaining configurations unchanged.

\subsection{Results of Medical Report Generation} \label{resultsMRG}

In this sub-section, we conduct extensive experiments on the Chinese/English report generation and compare it with current state-of-the-art report generators.

\subsubsection{Chinese Report Generation} 

As shown in Table~\ref{tab:PCCXray_results}, the baseline R2Gen+MAE achieves $0.660, 0.588, 0.536, 0.498, 0.594$ on our newly collected PCC-Xray dataset on the BLEU-1, BLEU-2, BLEU-3, BLEU-4, and ROUGE-L metrics, meanwhile, the results can be improved to $0.679, 0.609, 0.560, 0.523, 0.611$ when using the ViT backbone network pre-trained on our X-Ray dataset using the masked auto-encoder framework. This comparison demonstrates that pre-training on large-scale X-ray images indeed helps feature representation learning more than on natural images. When the context-aware masking strategy is adopted in the pre-training, the results can be further improved to $0.719, 0.656, 0.611, 0.577, 0.651$. It is easy to find that the context-aware masking works for our X-ray based pre-training foundation model.

\begin{table} 
\centering
\resizebox{0.48\textwidth}{!}{
\begin{tabular}{l|ccccc}
\hline \toprule [0.5 pt] 
\textbf{Method}         & \textbf{BLEU-1} & \textbf{BLEU-2} & \textbf{BLEU-3} & \textbf{BLEU-4}  & \textbf{ROUGE\_L} \\ 
\hline 
R2Gen+MAE         &0.660   &0.588    &0.536    &0.498   &0.594   \\ 
Ours              &0.697   &0.631    &0.583    &0.548    &0.635  \\
Ours ($w/o$ CaM)  &0.694   &0.627    &0.579    &0.543    &0.628  \\
\hline \toprule [0.5 pt] 
\end{tabular}} 
\caption{ Experimental results on the PCC-Xray dataset. $w/o$ denotes without the following item.}   
\label{tab:PCCXray_results}  
\end{table}

\subsubsection{English Report Generation}
As shown in Table~\ref{tab:IUXray_Results}, we also adapt our framework to handle the English medical report generation task. In this section, we report the experimental results on the IU-Xray dataset which is 0.677, 0.185, 0.395, 0.197 on the CIDEr, BLEU-4, ROUGE-L, METEOR metric, respectively. Compared with our baseline method R2Gen~\cite{chen2020R2Gen} which obtains 0.398, 0.165, 0.371, 0.187 on these metrics, our model achieves a significant improvement. Compared with other models, such as DCL~\cite{li2023dynamic} and MGSK~\cite{yang2022knowledge}, our results are also better than theirs. These results fully validated the effectiveness of our proposed foundation model for the perception of X-ray images.

\begin{table} 
\centering
\begin{tabular}{l|cccc}
\hline \toprule [0.5 pt] 
\textbf{Methods} & \textbf{CIDEr} & \textbf{BLEU-4} & \textbf{ROUGE-L} & \multicolumn{1}{l}{\textbf{METEOR}} \\ \hline
R2Gen~\cite{chen2020R2Gen} & 0.398 & 0.165 & 0.371 & 0.187 \\
KERP~\cite{li2019knowledge} & 0.280 & 0.162 & 0.339 & - \\
HRGP~\cite{li2018hybrid} & 0.343 & 0.151 & 0.322 & - \\
MKG~\cite{zhang2020radiology} & 0.304 & 0.147 & 0.367 & - \\
PPKED~\cite{liu2021exploring} & 0.351 & 0.168 & 0.376 & 0.190 \\
MGSK~\cite{yang2022knowledge} & 0.382 & 0.178 & 0.381 & - \\
CA~\cite{liu2021contrastive} & - & 0.169 & 0.381 & 0.193 \\
CMCL~\cite{liu2021competence} & - & 0.162 & 0.378 & 0.186 \\
DCL~\cite{li2023dynamic} & 0.586 & 0.163 & 0.383 & 0.193 \\ \hline
Ours & \textbf{0.677} & \textbf{0.185} & \textbf{0.395} & \textbf{0.197} \\ 
Ours ($w/o$ CaM) & 0.611 & 0.171 & 0.375 & 0.195 \\ 
\hline \toprule [0.5 pt] 
\end{tabular}
\caption{The performances of our proposed model compared with other state-of-the-art systems on IU-Xray dataset. The
best results in each column are highlighted in bold. $w/o$ denotes without the following item.} 
\label{tab:IUXray_Results}
\end{table}

\subsection{Results of Diseases Recognition} \label{resultsDisR} 
As shown in Table~\ref{tab:disease_pred}, we report our experimental results on the disease prediction task and compare them with recent state-of-the-art recognition models. Obviously, our recognition results are comparable to these models but still inferior to them. We think this may be caused by the fact that our model is pre-trained on high-resolution X-ray images, but the images in RSNA-Pneumonia dataset~\cite{shih2019RSNAPneumonia} are the standard resolution. In our future works, we will consider pre-training on multi-scale X-ray images to further improve the generation and robustness of our model. Another possible reason is that the VTB is proposed for pedestrian attribute recognition, the parameter configurations may be different from the disease recognition task.

\begin{table} 
\centering
\begin{tabular}{l|l|ccc}
\hline \toprule [0.5 pt] 
\multirow{2}{*}{\textbf{Backbone}} & \multirow{2}{*}{\textbf{Method}} & \multicolumn{3}{c}{\textbf{RSNA-Pneumonia (AUC)}} \\ \cline{3-5}  
 &  & 1\% & 10\% & 100\% \\ \hline
\multirow{4}{*}{CNN} 
 &GLoRIA~\cite{huang2021gloria} & 86.1 & 88.0 & 88.6 \\
 &PRIOR~\cite{cheng2023prior} & 85.7 & 87.1 & 89.2 \\
 &MedKLIP\#~\cite{wu2023medklip} & 87.3 & 88.0 & 89.3 \\
 &KAD~\cite{zhang2023knowledge} & 89.8 & 91.8 & 92.5 \\ \hline
\multirow{9}{*}{Transformer} 
 &MAE~\cite{he2022MAE} & 84.2 & 89.6 & 91.3 \\
 &REFERS~\cite{zhou2022generalized} & 89.4 & 91.6 & 92.7 \\
 &MGCA\#~\cite{wang2022multi} & 90.7 & 92.6 & 93.4 \\
 &MRM~\cite{zhou2023advancing} & 91.3 & 92.7 & 93.3 \\
 &ECAMP~\cite{wang2023ecamp} & 91.5 & 92.9 & 93.8 \\ \cline{2-5} 
 & Ours & 83.4 & 86.3 & 88.2 \\ 
 & Ours ($w/o$ CaM) & 46.3 & 83.7 & 86.9 \\
\hline \toprule [0.5 pt] 
\end{tabular}
\caption{Results of disease recognition on RSNA-Pneumonia dataset. Methods with \# leverage disease-level annotations. $w/o$ denotes without the following item.}
\label{tab:disease_pred}
\end{table}

\subsection{Ablation Study} \label{ablationStudy} 

In this sub-section, we conduct extensive experiments to further help the readers better understand our framework.

\subsubsection{Random Masking \textit{vs} Context-aware Masking} 
To verify the effectiveness of our proposed Context-aware Masking (CaM, for short) for the X-ray based masked auto-encoder, we compare the performance of report generation with and without ($w/o$) the CaM, as shown in Table~\ref{tab:disease_pred}. With the help of CaM, we achieve 83.4, 86.3, 88.2 when using 1\%, 10\%, and 100\% of the training data of RSNA dataset, but we only get 46.3, 83.7, and 86.9 on these settings when removing this module, i.e., Ours ($w/o$ CaM). This comparison fully validated the effectiveness and importance of the context-aware masking strategy. 
Similar conclusions can also be drawn from Table~\ref{tab:PCCXray_results} and Table~\ref{tab:IUXray_Results}.

\subsubsection{Does High Definition X-ray Image Works for Report Generation?} 
As shown in Table~\ref{tab:IUXray_diffSizes}, we set different resolutions of X-ray images to test its influence on the final results, including $384 \times 384$, $448 \times 448$, and $512 \times 512$. We can find that better results can be obtained when the resolution is set as $384 \times 384$, i.e., 0.185, 0.197, 0.395, 0.677 on the BLEU-4, METEOR, ROUGE-L, CIDEr metric. This result is consistent with current vision models which can achieve higher performance when slightly increasing the resolution from $224 \times 224$. However, the performance drops when further increasing the resolution, as reported in~\cite{liu2024vmamba, zhu2024visionmamba}.

\begin{table} 
\centering
\begin{tabular}{l|cccc}
\hline \toprule [0.5 pt] 
\textbf{Resolution} & \textbf{BLEU-4} & \textbf{METEOR} & \textbf{ROUGE-L} & \textbf{CIDEr} \\ 
\hline
$384 \times 384$ & \textbf{0.185} & \textbf{0.197} & \textbf{0.395} & \textbf{0.677} \\
$448 \times 448$ & 0.162 & 0.197 & 0.369 & 0.625 \\
$512 \times 512$ & 0.163 & 0.192 & 0.361 & 0.596 \\ 
\hline \toprule [0.5 pt] 
\end{tabular}
\caption{ Performance on the IU-Xray test dataset using different input sizes. }
\label{tab:IUXray_diffSizes}
\end{table}

\subsubsection{The Curve of Relationship between Epoch and Accuracy} 
\begin{figure}
\centering
\includegraphics[width=1\linewidth]{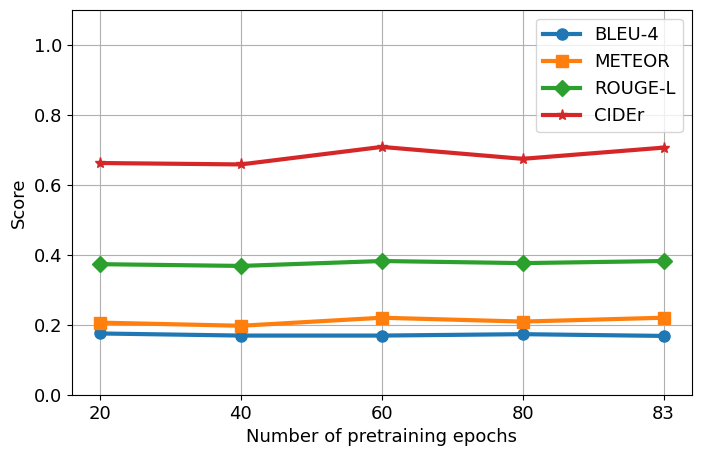}
\caption{ Variation of the accuracy on the IU-Xray testing dataset in the training phase. }
\label{fig:EpochAccuracy}
\end{figure}

As shown in Table~\ref{tab:EpochAccuracy} and Fig.~\ref{fig:EpochAccuracy}, we report the corresponding results on the IU-Xray testing subset in the pre-training phase, i.e., $20^{th}$, $40^{th}$, $60^{th}$, $80^{th}$, and $83^{th}$ epoch. Generally speaking, better results can be obtained in the late stage of our pre-training.

\begin{table}
\centering
\begin{tabular}{l|cccc}
\hline \toprule [0.5 pt] 
\textbf{Method} & \textbf{BLEU-4} & \textbf{METEOR} & \textbf{ROUGE-L} & \textbf{CIDEr} \\ \hline
Epoch-20 & \textbf{0.175} & 0.206 & 0.373 & 0.662 \\
Epoch-40 & 0.169 & 0.197 & 0.368 & 0.658 \\
Epoch-60 & 0.169 & 0.220 & 0.382 & \textbf{0.708} \\
Epoch-80 & 0.173 & 0.209 & 0.376 & 0.674 \\
Epoch-83 & 0.168 & \textbf{0.220} & \textbf{0.382} & 0.706 \\ 
\hline \toprule [0.5 pt] 
\end{tabular}
\caption{ The detailed accuracy on the IU-Xray testing dataset in the training phase. }
\label{tab:EpochAccuracy}
\end{table}

\begin{figure}
    \centering
    \includegraphics[width=1\linewidth]{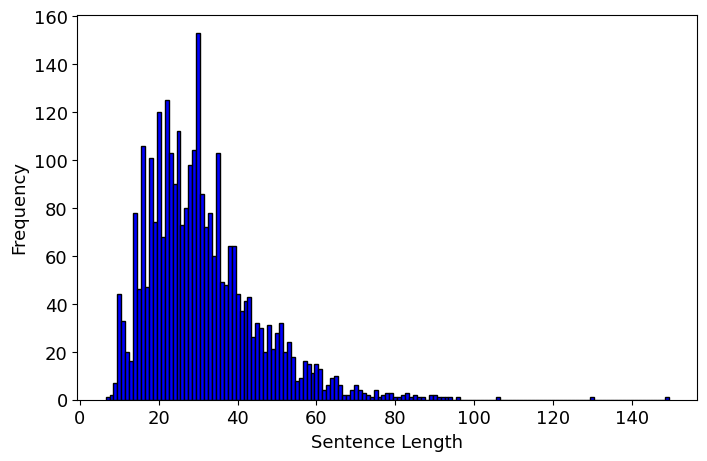}
    \caption{Distribution of sentence length of IU-Xray dataset.}
    \label{fig:seq_frequency}
\end{figure}

\begin{figure} 
    \centering
    \includegraphics[width=1\linewidth]{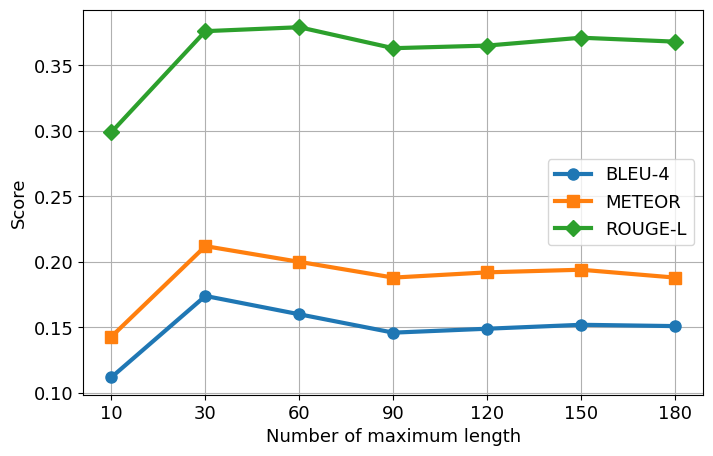}
    \caption{ Variation of different maximum length of report generator on the IU-Xray testing subset.} 
    \label{fig:max_seq_length}
\end{figure}

\begin{figure*}[]  
\captionsetup[subfigure]{labelformat=empty}
\resizebox{\textwidth}{!}{
    \centering
        \begin{subfigure}[b]{0.16\textwidth}  
            \caption{Origin} 
            \includegraphics[width=\textwidth]{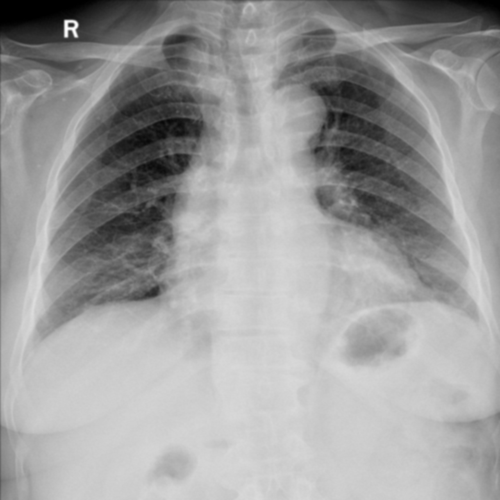}
        \end{subfigure}
        
        \begin{subfigure}[b]{0.16\textwidth} 
            \caption{Mask}
            \includegraphics[width=\textwidth]{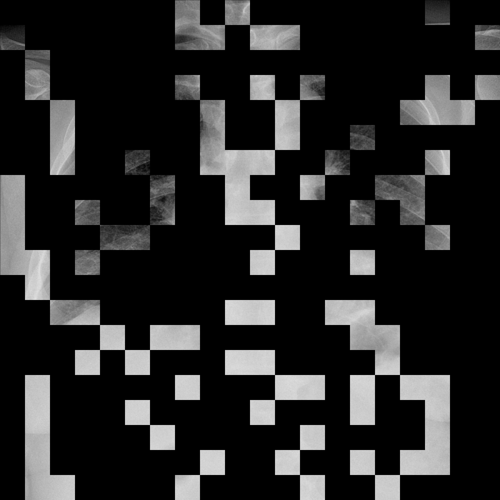}
        \end{subfigure}
        \begin{subfigure}[b]{0.16\textwidth} 
            \caption{Reconstruction}
            \includegraphics[width=\textwidth]{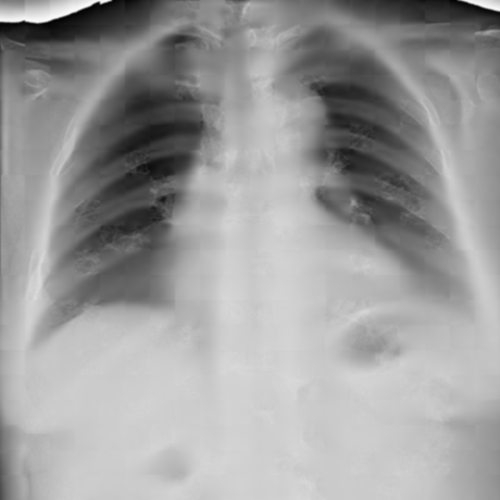}
        \end{subfigure} 
    \hspace{0.2cm}
        \centering 
        \begin{subfigure}[b]{0.16\textwidth}
            \caption{Origin}
            \includegraphics[width=\textwidth]{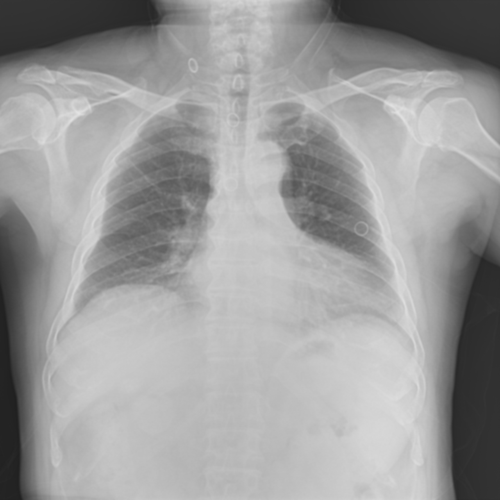}
        \end{subfigure}
        \begin{subfigure}[b]{0.16\textwidth} 
            \caption{Mask}
            \includegraphics[width=\textwidth]{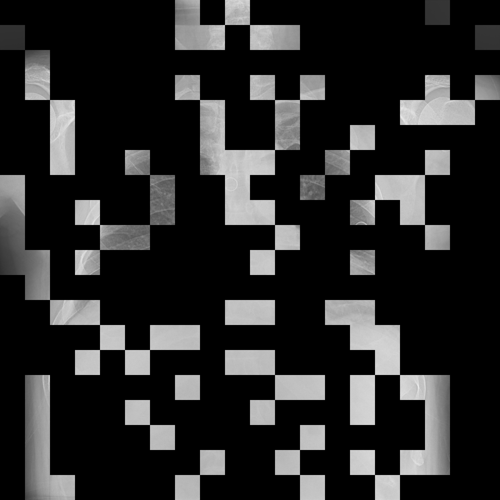} 
        \end{subfigure}
        \begin{subfigure}[b]{0.16\textwidth} 
            \caption{Reconstruction}
            \includegraphics[width=\textwidth]{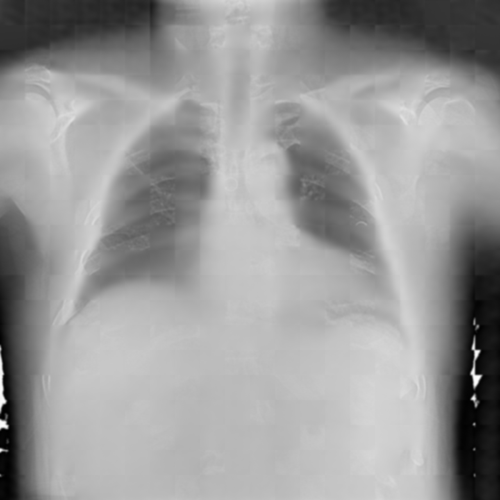}
        \end{subfigure} 
}
\resizebox{\textwidth}{!}{
        \centering 
        \includegraphics[width=0.16\textwidth]{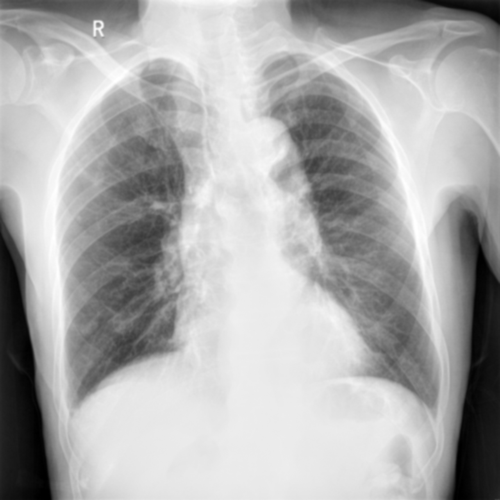}
        \includegraphics[width=0.16\textwidth]{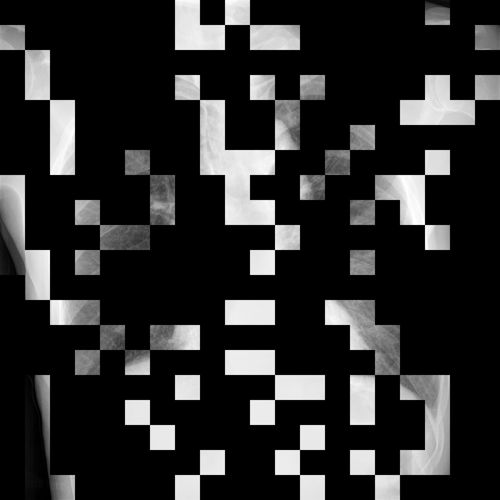} 
        \includegraphics[width=0.16\textwidth]{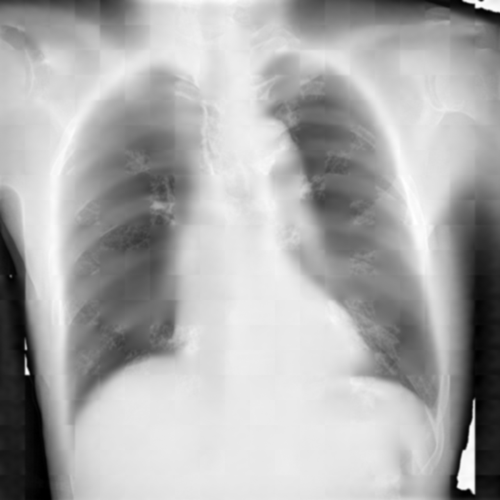}
    \hspace{0.2cm}
        \centering 
        \includegraphics[width=0.16\textwidth]{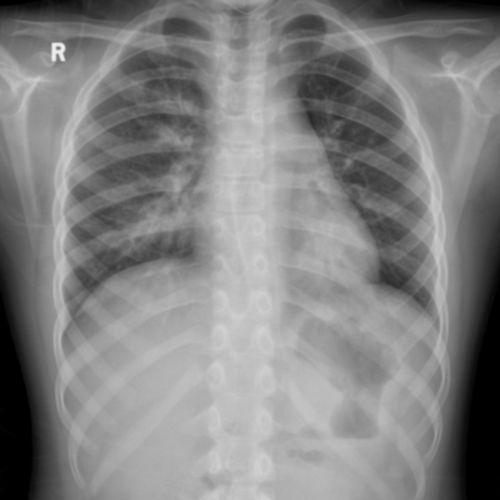}
        \includegraphics[width=0.16\textwidth]{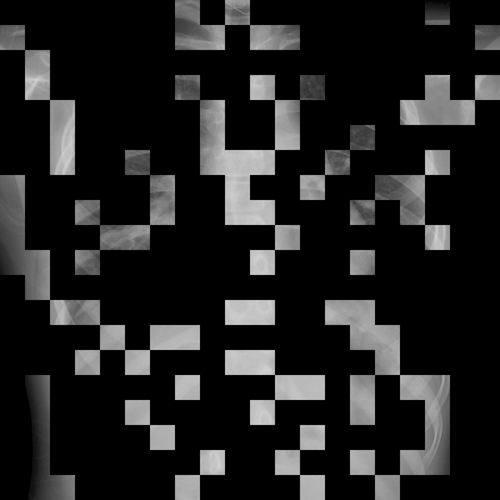} 
        \includegraphics[width=0.16\textwidth]{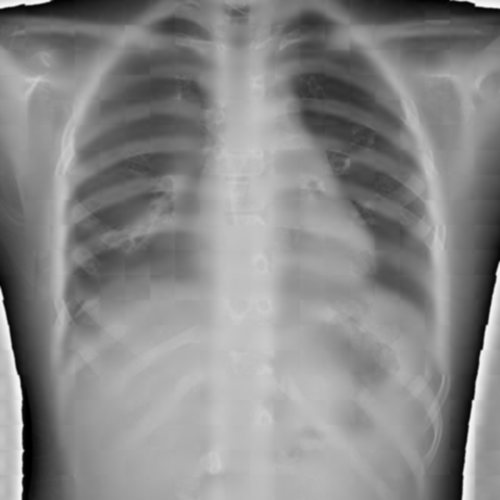}
    }
    \caption{ Visualization of the reconstructed masked tokens on our newly collected PCC-Xray dataset.}  
    \label{fig:mask_region}  
\end{figure*}

\begin{figure*}
\captionsetup[subfigure]{labelformat=empty}
    \resizebox{\textwidth}{!}{
            \centering  
            \begin{subfigure}[h]{0.16\textwidth} 
            \caption{Origin}
                \includegraphics[width=\textwidth]{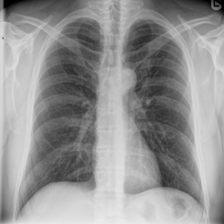}
            \end{subfigure}
            \begin{subfigure}[h]{0.16\textwidth} 
            \caption{Lungs}
                \includegraphics[width=\textwidth]{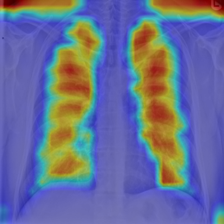}
            \end{subfigure}
        \hspace{0.2cm}
            \centering  
            \begin{subfigure}[h]{0.16\textwidth} 
            \caption{Origin}
                \includegraphics[width=\textwidth]{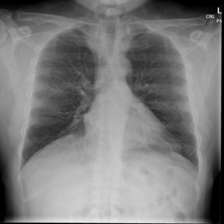}
            \end{subfigure}
            \begin{subfigure}[h]{0.16\textwidth} 
            \caption{Lungs}
                \includegraphics[width=\textwidth]{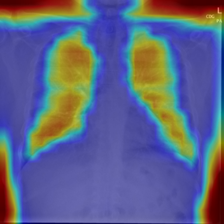}
            \end{subfigure}
        \hspace{0.2cm}
            \centering  
            \begin{subfigure}[h]{0.16\textwidth} 
            \caption{Origin}
                \includegraphics[width=\textwidth]{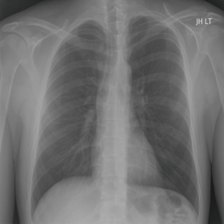}
            \end{subfigure}
            \begin{subfigure}[h]{0.16\textwidth} 
            \caption{Lungs}
                \includegraphics[width=\textwidth]{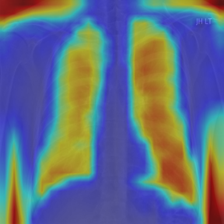}
            \end{subfigure}
    }
\vspace{0.2cm}
    \resizebox{\textwidth}{!}{
    \centering
        \centering  
        \includegraphics[width=0.16\textwidth]{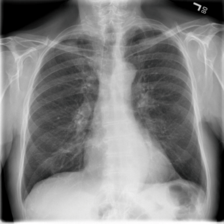}
        \includegraphics[width=0.16\textwidth]{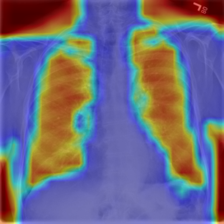}
    \hspace{0.2cm}
        \centering  
        \includegraphics[width=0.16\textwidth]{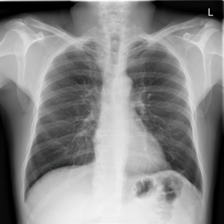}
        \includegraphics[width=0.16\textwidth]{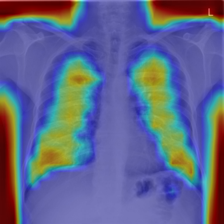}
    \hspace{0.2cm}
        \centering  
        \includegraphics[width=0.16\textwidth]{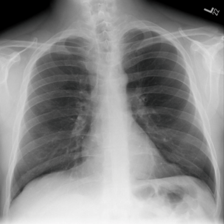}
        \includegraphics[width=0.16\textwidth]{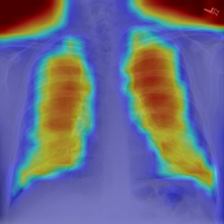}
    }
    \caption{Visualizations of activation response maps on the IU-Xray dataset.}  
    \label{fig:iu_atten_visual}  
\end{figure*}

\begin{figure*}
\centering
\resizebox{\textwidth}{!}{
\begin{tabular}{lp{9cm}p{9cm}}
\hline \toprule [0.5 pt] 
\multicolumn{1}{c}{\textbf{Image}} & \multicolumn{1}{c}{\textbf{Ground Truth}} & \multicolumn{1}{c}{\textbf{Ours}} \\ 
\hline \toprule [0.5 pt]  
\begin{minipage}[b]{0.3\columnwidth}\vspace{6pt}\centering\raisebox{-.8\height}{\includegraphics[width=\linewidth]{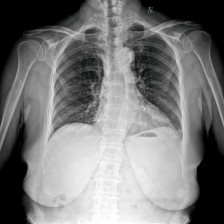}}\end{minipage} & \begin{CJK*}{UTF8}{gbsn}  胸廓对称，气管居中；两侧肋骨、肋间隙正常；两下肺纹理增多增粗，其间见斑点、小斑片状密度增高影，两侧肺门和纵隔影未见明显异常；主动脉结突出伴有钙化影，心影横径稍增大；膈肌平滑，双侧肋膈角锐利。\end{CJK*} & \begin{CJK*}{UTF8}{gbsn}  两胸廓对称，肋骨走形自然，未见畸形及骨质破坏。双肺纹理增多、增粗、紊乱。两肺未见明显实质性病变。心影横径增宽。双膈面尚光整，肋膈角尚锐利。\end{CJK*} \\
\begin{minipage}[b]{0.3\columnwidth}\vspace{6pt}\centering\raisebox{-.6\height}{\includegraphics[width=\linewidth]{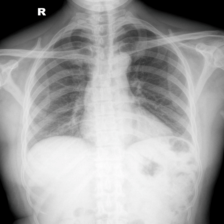}}\end{minipage} & \begin{CJK*}{UTF8}{gbsn}  两侧胸廓对称，两肺未见明显实质性病变，两侧膈面光滑，两侧肋膈角锐利。心影形态、大小未见明显异常。\end{CJK*} & \begin{CJK*}{UTF8}{gbsn}  两侧胸廓对称，两肺未见明显实质性病变，两侧膈面光滑，两侧肋膈角锐利。心影形态、大小未见明显异常。\end{CJK*} \\
\begin{minipage}[b]{0.3\columnwidth}\vspace{6pt}\centering\raisebox{-.6\height}{\includegraphics[width=\linewidth]{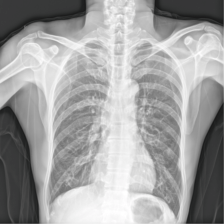}}\end{minipage} \vspace{6pt}& \begin{CJK*}{UTF8}{gbsn}  两肺纹理稍增多，可见散在分布斑点状高密度影，边界尚清，心影大小形态大致正常范围内，双侧膈肌光滑，肋膈角锐利。\end{CJK*} & \begin{CJK*}{UTF8}{gbsn}  两肺纹理增多，右上肺野见斑点、条状、结节状中等密度影，边界部分清晰。心影大小形态大致正常范围内，双侧膈肌光滑，肋膈角锐利。\end{CJK*} \\ 
\hline \toprule [0.5 pt]  
\end{tabular}
}
\caption{The X-ray image based medical report generation on our PCC-Xray dataset.}
\label{fig:cn_report}
\end{figure*}

\subsubsection{Influence on the Maximum Length of Medical Report Predicted by the Report Generator}  
As shown in Fig.~\ref{fig:seq_frequency}, we give a visualization of the distribution of the number of words in each sentence on the IU-Xray dataset. We can find that most sentences contain about 20-40 words. Interestingly, as the results reported in Fig.~\ref{fig:max_seq_length}, the peak results can be achieved when the maximum length of the medical report predicted by the report generator is set as 30. Therefore, we choose this hyper-parameter as 30 in this work for the IU-Xray dataset.

\subsection{Visualization}  \label{visualization}
In this sub-section, we give some visualizations to help the readers better understand the effectiveness of our model, including the reconstructed masked tokens (Section~\ref{sec:ReconstMaskedTokens}), the activation response maps (Section~\ref{sec:ActResponseMaps}), generated medical reports (Section~\ref{sec:MedicalReportGen}), and predicted diseases (Section~\ref{sec:DiseasePred}).

\subsubsection{Reconstructed Masked Tokens} \label{sec:ReconstMaskedTokens} 
As shown in Fig.~\ref{fig:mask_region}, we provide some representative samples predicted by our model. The $1^{th}$ and $4^{th}$ column are the raw X-ray images, the $2^{th}$ and $5^{th}$ column are masked images, and the $3^{th}$ and $6^{th}$ column are the reconstructed images. We can find that our proposed context-aware masking strategy guided MAE framework predict the masked tokens well.

\subsubsection{Activation Response Maps}  \label{sec:ActResponseMaps} 
As shown in Fig.~\ref{fig:iu_atten_visual}, given the text \textit{lungs}, we can find that the activation maps can accurately highlight the target regions. Therefore, we can achieve a higher performance on the downstream tasks. However, the activation maps are imperfect, as the background regions are also highlighted.

\subsubsection{Medical Report Generation} \label{sec:MedicalReportGen}
In addition to aforementioned visualization on the reconstructed masked tokens and activation response maps, we also show the generated medical reports on the PCC-Xray dataset, as shown in Fig.~\ref{fig:cn_report}. It is easy to find that our model performs well and accurately predicts the reports.

\subsubsection{Disease Prediction}  \label{sec:DiseasePred} 
\begin{figure}
\captionsetup[subfigure]{labelformat=empty}
                \centering  
            \begin{subfigure}[h]{0.2\textwidth} 
            \caption{Not Lung Opacity}
                \includegraphics[width=\textwidth]{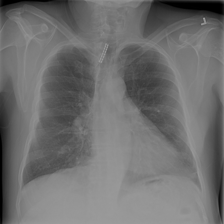}
            \end{subfigure}
            \begin{subfigure}[h]{0.2\textwidth} 
             \caption{Lung Opacity}
                \includegraphics[width=\textwidth]{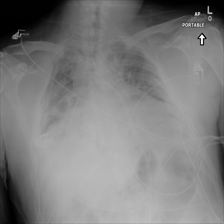}
            \end{subfigure}
    \caption{Visualizations of lung opacity prediction in RSNA-Pneumonia dataset.}  
    \label{fig:VIS_diseasePred}
\end{figure}

As shown in Fig.~\ref{fig:VIS_diseasePred}, given the X-ray image from the RSNA-Pneumonia dataset and all the labels (binary classification) we need to recognize, our model can predict the disease accurately.

\subsection{Limitation Analysis} \label{limitation}
This work attempts to conduct self-supervised pre-training based on MAE (Masked Auto-Encoder) on a high-resolution X-ray dataset and validates it on two mainstream medical downstream tasks. The results indicate that our X-ray based model indeed achieves promising results. Experimental results fully demonstrate that Transformer-based big model frameworks can ensure decent results, but they still cannot achieve the astonishing performance boost seen in large language models. We think the current model can still be improved from the following perspectives: 
1). The current framework adopts the Transformer as the core block, bringing a huge computation cost in the pre-training phase. 
2). Only X-ray images are used in the pre-training phase which ignores the semantic cues, therefore, the overall performance may still sub-optimal. 
3). Current mainstream backbone networks adopts $224 \times 224$ as their default resolution of the input image, however, the specific design to address the high-resolution images still further exploring.

\section{Conclusion and Future Works} \label{sec:conclusion}
In this work, we summarize the issues of existing X-ray image based pre-training methods and propose to pre-training a high-definition foundation model. Specifically, we follow the self-supervised pre-training framework masked auto-encoder (MAE) and design a new context-aware masking strategy. For the downstream tasks, we test our model on both English/Chinese report generation and disease prediction. The experiments on multiple benchmark datasets fully validated the effectiveness of our model.

In future work, further improvements are still needed to pursue breakthroughs. Specifically speaking, 
1). The X-ray based vision foundation model proposed in this paper is based on Transformer but has a complexity of $\mathcal{O}(N^2)$, resulting in high memory consumption and computational costs when handling high-resolution X-ray data. In the future, we will attempt to introduce new lightweight network architectures (such as State Space Model/Mamba~\cite{wang2024SSMSurvey}) to address its computational complexity issues. 2). Pre-training purely from a visual self-supervised manner can yield decent improvements, but the overall accuracy is still not satisfactory. Subsequently, we will consider multi-modal pre-training approaches, incorporating large language models, knowledge graphs, etc., to further enhance the representation ability of the visual foundation model. 3). In addition to pre-training, to enhance performance on downstream tasks (such as medical reports), we will explore the introduction of knowledge graphs or other useful prompts to improve its performance in text generation.

\label{DiscSec}
\bibliographystyle{IEEEtran}
\bibliography{reference}

\end{document}